# Spatially extended nature of resistive switching in perovskite oxide thin films

Xin Chen*, NaiJuan Wu, John Strozier and Alex Ignatiev

*Texas Center for Advanced materials,*

*University of Houston, Houston, TX 77204-5004*

*Abstract:* -We report the direct observation of the electric pulse induced resistance-change (EPIR) effect at the nano scale on $La_{1-x}Sr_xMnO_3$ (LSMO) thin films by the current measurement AFM technique. After a switching voltage of one polarity is applied across the sample by the AFM tip, the conductivity in a local nanometer region around the AFM tip is increased, and after a switching voltage of the opposite polarity is applied, the local conductivity is reduced. This reversible resistance switching effect is observed under both continuous and short pulse voltage switching conditions. It is important for future nanoscale non-volatile memory device applications.

[*] Email: xinchen@svec.uh.edu; Fax: 713-747-7724

Advanced non-volatile memory device research has recently drawn broad interest as the result of new materials systems being studied for such applications[1] including colossal magnetoresistance (CMR) materials[2]. The recent activity in the electrical pulse induced resistance-change (EPIR) switching effect in CMR perovskite oxides[3] is one area which not only shows promise for overcoming the shortcomings in the current semiconductor non-volatile memory technologies[1,4], but also brings questions as to the extent and basis for the resistance switching phenomenon. For instance, it has been suggested that resistance switching inside a material by a pulse of one polarity and reversal by a pulse of the opposite polarity appears to violate parity conservation.[5] As a result, it has been suggested the EPIR effect must be a contact surface effect, which may not involve the CMR material. Here we report by the direct measurement of current in the atomic force microscopy (I-AFM)[6] that the EPIR effect in a $La_{1-x}Sr_xMnO_3$ (LSMO) thin film extends over a region of the LSMO perovskite oxide material near the interface of metal electrode and LSMO thin film. These observations are very important for future non-volatile memory device applications[1,4].

The sample was prepared by pulsed laser deposition (PLD) of an LSMO thin film onto an iridium (Ir) film bottom-electrode layer, which had been grown on a $TiN/SiO_2/Si$ wafer substrate. The LSMO thin film deposition was carried out at 550°C under 150mtorr oxygen ambient, and a portion of the Ir layer was masked to allow for lead wire connection. The surface microstructure of the LSMO film and the resistance switching behavior of LSMO device were examined by AFM and I-AFM measurements with a PSIA XE-100 scanning probe microscope system.

Fig. 1a schematically illustrates the set-up of I-AFM measurement. A metallic coated Si AFM tip is put in contact with the sample surface, and a bias voltage is applied between the tip and the sample. The current flow due to the applied bias can be monitored both at a certain point on the surface, and in a two-dimensional scan. The system is connected to an external current amplifier for increased current sensitivity. AFM and I-AFM images for the LSMO film surface are presented in Figs. 1b and 1c along with their line scan plots identified by solid lines marked on the respective images. XRD analysis indicates that the LSMO film grown on the Ir-Si substrate is polycrystalline. AFM analysis shows that the LSMO film surface has nanometer scale smoothness (RMS roughness of 1-2nm), with a few of submicron grains of around 5nm height, which are often seen on film surface fabricated by PLD method. A submicron grain has been used as a reference to correct tip mis-location effect due to the thermal drift of instrument. The I-AFM 2-D scan image of Fig. 1c obtained simultaneously with the AFM scan of Fig. 1b indicates that the conduction through the LSMO film is related to the thin film surface structure and has a granular distribution that correlates with the fine structured nano islands on the sample surface. The marked dashed frame in Fig. 1c is the area where we will present detailed EPIR resistance switch data for the sample.

Before further I-AFM examination, Ag top-electrode pads with ~200 μm diameter were deposited on the LSMO/Ir/TiN/SiO$_2$/Si sample by DC sputtering. Fig. 2 presents the EPIR[3] switching behavior for the sample observed by applying ~±3V, 200ns pulses across an Ag top-contact and the Ir bottom-contact on the sample. A positive pulse of 3.2V decreases the sample resistance to its low resistance state (LRS) of about 850Ω; and a negative pulse of −3V, increases the sample resistance to the high resistance state (HRS) of about 1150Ω. This result confirms that the LSMO perovskite thin film is exhibiting the EPIR switching behavior.[3]

I-V measurements were then taken using the AFM tip with a continuous bias voltage applied between the tip and the Ir bottom-electrode. This resulted in the I-V hysteresis curve shown in Fig. 3a. The arrows indicate the voltage scan direction. The I-V curve is non-linear, and at positive voltage, exhibits sample switching to a lower resistance state, while at negative voltage, to a higher resistance state. The application of continuous voltage instead of a pulse voltage results in partial switching of the sample at a voltage much lower than presented in Fig. 2.

The sample and the conductive AFM tip were then configured such that they could be connected to an external pulsing circuit. Sample resistance switching was studied by applying pulses to the sample through the I-AFM tip. 2-D I-AFM scans were then taken over the framed area in Fig. 1c after the pulse application through the AFM tip. An arrow indicates on the I-AFM scan of Fig. 3b, the location where the LSMO thin film was switched to the HRS by the application of negative voltage pulse through the tip. The dark area at the arrow tip indicates low current flow, and hence high resistance at that location. The scan in Fig. 3c is of the same location after it was switched to the LRS by application of a positive voltage pulse through the AFM tip. The area of the sample near the pulse position shows a high intensity, and hence a lower resistance. Repeat of this HRS-LRS switching is shown in Figs. 3d and 3e. Current line scans through the region of the sample where the switching pulse was applied are also shown in Figs. 3b and 3c. The line scans directly show local conductivity change in the region near where the pulse was applied through the tip as observed by the large increase in current in the 40nm to 160nm region of the scan of Fig. 3c after switching to the LRS. Although the tip radius was ~30nm, tip contact area is expected to be much smaller due to its convex shape. This is partially reflected by the good lateral resolution in the I-AFM images. The observed resistance switch region of the scan, however is ~ 100nm or larger in extent, indicting that the switch region is not limited to the tip/LSMO film contact surface, but extends significantly into the LSMO film.

Previous studies had diverged on bulk[7] or contact surface switching[5] contributions to the EPIR effect. Our recent work[8,9,10] has suggested that the EPIR switching characteristics are attributed to the oxide material of the EPIR sandwich sample, and hence the switching effect extends significantly beyond the contact interface region. The present nano scale I-AFM analysis further proves that the EPIR switching effect indeed extends over a region of the order of ~100nm laterally around the contact surface, and rules out the possibility that EPIR is only a contact surface effect. Since the voltage was applied vertically across the sample, it is expected that the EPIR switching effect would occur in a similar extension vertically. It is well to note again that the resistance switching effect is not limited to the contact surface between the electrode and the perovskite oxide, but penetrates some extent into the oxide material.

Conduction in manganite oxides at room temperature is considered to occur by hoping of carriers consisting of small polaron along –Mn-O-Mn-O- chain and/or by conducting phase droplets in a paramagnetic insulating matrix. In either case, the character of the carriers is thought to be holes. It has also been reported that small changes in oxygen concentration in CMR materials results in large resistance changes[11]. In addition, current injection of electrons can cause oxygen migration in a perovskite oxide[12]. Considering that oxygen deficiency generally exists in perovskite oxides, electric pulse current driven ion motion is being presented here as active agent for the resistance switching: moving oxygen vacancies toward (ions away from) the electrode contact interface during a negative pulse, which decreases the electron wave function overlap between manganese and oxygen ions, i.e., localizes or traps the electrons in the interface region, thereby increasing the resistance. And vise-versa for a positive pulse, where electrons are delocalized and the resistance is decreased. We need to stress that due to the strong electron correlated nature of CMR materials, small changes in oxygen concentration may result in large resistance changes in CMR material[11]. The distance of motion of the ions/vacancies in the EPIR effect is reflected by the spatial extent of the resistivity change in the I-AFM measurements - over ~100nm during a nominal ~600ns pulse time. Such data indicates that the mobility of ions/vacancies in the proposed current enhanced diffusion is more than $10^5$ times greater than nominal thermal diffusion of oxygen ions in normal oxides[10,12,13].

Furthermore, the oxygen stoichiometry change induced by electric pulses might not be uniformly distributed within the interface region. In particular, phase separation and phase transitions are known to exist in the strong electron correlated CMR material systems.[2] In such systems, electron localization and de-localization can result in micro-nano texturing in the material, which could produce nanoscale metallic and insulating phases in the material[14]. The balance between charge-localized insulating phases and charge-delocalized conducting phases may lead to a number of "glassy features" in the CMR material at room temperature, that could demonstrate hysteresis and memory effects.[15] As seen in Figure 3, the resistance switching region is not a uniform round shape, but shows nano textured granular conductivity distributions. Additional details of the oxygen motion relation with resistance change will be published elsewhere.

In summary, the electrical pulse induced reversible resistance switching effect in perovskite oxides is shown to occur over an extended region of the active oxide thin film: about 40-160 nm around the contact interface into the CMR material, under both continuous voltage switching and short pulse voltage switching conditions. This clearly identifies the perovskite oxide material as playing the major role in the resistive switching process, and the EPIR effect is an intrinsic property of CMR oxide material around the metal-oxide contact region. The electric pulse driven oxygen ion/vacancy motion as an active agent can be used to explain the resistance switching mechanism for the EPIR effect. Furthermore, it is also shown that resistive switching can be accomplished in CMR oxides on the nano scale at room temperature. Such nano region switch phenomenon might be used to fabricate memory devices with density up to $10^{10}$ bits/cm$^2$, boding well for the future fast, high density resistive random access memory development based on the non-volatile resistance change effect.

**Acknowledgements.** We acknowledge the assistance of Y. Q. Wang and Y. B. Nian in this effort. Partial support of Sharp Laboratories of America, NASA, the R. A. Welch Foundation, and the State of Texas through the Texas Center for Advanced Materials is greatly acknowledged.

[1.] C. U. Pinnow, and T. Mikolajick, J. Electrochem. Soc. **151(6),** K13-K19, (2004).
[2.] E. Dagotto, Science **309,** 257-262 (2005).

3. S. Q. Liu, N. J. Wu, and A. Ignatiev, Appl. Phys. Lett. **76(19),** 2749-2751 (2000).

4. Y. Tokura, Physics Today, **56(7),** 50-55 (2003).

5. A. Baikalov, Y. Q. Wang, B. Shen, B. Lorenz, S. Tsui, Y. Y. Sun, Y.Y. Xue, and C. W. Chu, Appl. Phys. Lett. **83(5),** 957-959, (2003).

6. J. M. Mao, I. K. Sou, J. B. Xu, and I. H. Wilson, J. Vac. Sci. Technol. B **16(1),** 14-18 (1998).

7. K. Aoyama, K. Waku, A. Asanuma, Y. Uesu, and T. Katsufuji, Appl. Phys. Lett. **85(7),** 1208-1210, (2004).

8. X. Chen, N. J. Wu, J. Strozier, and A. Ignatiev, Appl. Phys. Lett. **87,** 233506 (2005).

9. C. Papagianni, Doctoral Dissertation, University of Houston (2005).

10. Y. Nian, J. Strozier, N. J. Wu, X. Chen, and A. Ignatiev to be published.

11. H. L. Ju, J. Gopalakrishnan, J. L. Peng, Q. Li, G. C. Xiong, T. Venkatesan, and R. L. Greene, Phys. Rev. B **51(9),** 6143-6146, (1995).

12. N. A. Tulina, and V. V. Sirotkin, Physica C **400(3-4),** 105-110, (2004).

13. A. Gramm, T. Zahner, U. Spreitzer, R. Rossler, J. D. Pedarnig, D. Bauerle, and H. Lengfellner, Europhys. Lett. **49(4)**, 501-506 (2000).

14. K. H. Ahn, T. Lookman, and A. R. Blshop, Nature, **428,** 401-404 (2004).

15. V. Markovich, E. S. Vlakhov, Y. Yuzhelevskii, B. Blagoev, K. A. Nenkov, and G. Gorodetsky, Phys. Rev. B **72,** 134414 (2005).

**List of Figures**

**Figure 1** a) A schematic diagram of the I-AFM apparatus, b) the AFM plot and c) the I-AFM plot of a specific region of the sample obtained on the LSMO thin film surface.

**Figure 2** EPIR switching under pulsing through contact pads on the LSMO/Ir/Si sample.

**Figure 3** a) The typical continuous I-V loop for the LSMO film using the I-AFM tip, b) I-AFM 2-D scan and line scan of a tip-switch of the high resistance state (HRS) region and c) low resistance state (LRS) region, and d) and e) are repeats of b) and c).

**Figure 1** X. Chen et al

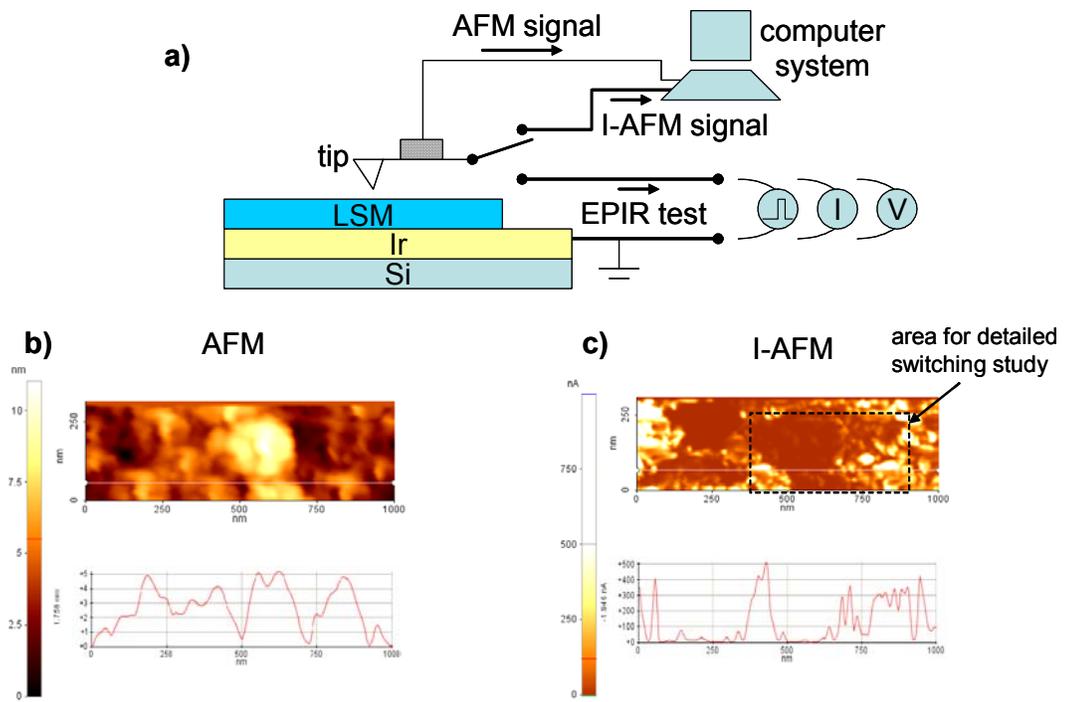

**Figure 2** X. Chen et al

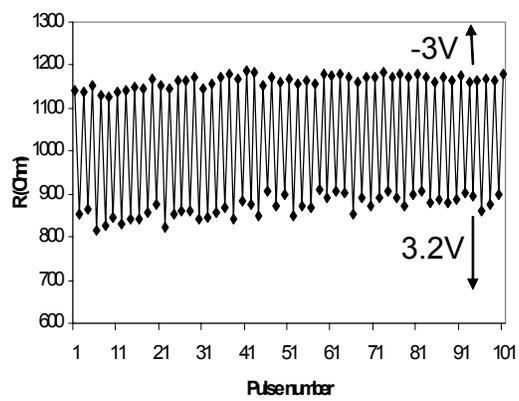

**Figure 3** X. Chen et al

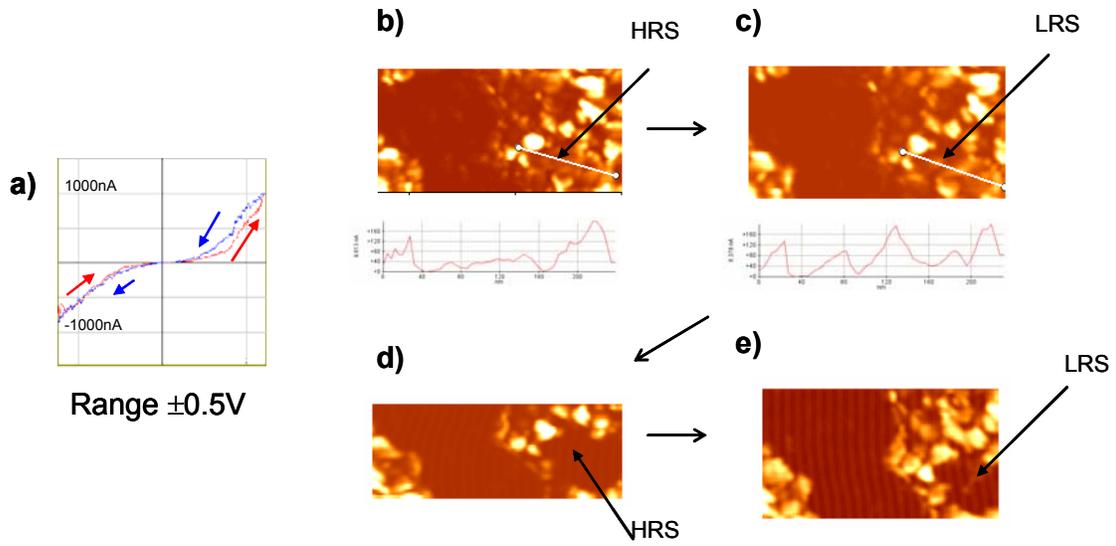